\documentclass[twocolumn,showkeys,showpacs,preprintnumbers,superscriptaddress,
amsmath,amssymb,nofootinbib]{revtex4}

\usepackage{graphicx}
\usepackage{epsfig}

\newcommand{\degr}{\mbox{$^\circ$}}

\newcommand{\po}{\mbox{$\pi^0$}}

\newcommand{\ee}{\mbox{$e^+e^-$}}

\newcommand{\heeg}{\mbox{$\et\rightarrow e^+e^-\gamma$}}
\newcommand{\heeee}{\mbox{$\et\rightarrow e^+e^-e^+e^-$}}
\newcommand{\hmmmm}{\mbox{$\et\rightarrow \mu^+\mu^-\mu^+\mu^-$}}
\newcommand{\hppmm}{\mbox{$\et\rightarrow \pi^+\pi^-\mu^+\mu^-$}}
\newcommand{\hmmg} {\mbox{$\et\rightarrow \mu^+\mu^-\gamma$}}
\newcommand{\hee}{\mbox{$\et\rightarrow e^+e^-$}}
\newcommand{\heemm} {\mbox{$\et\rightarrow e^+e^-\mu^+\mu^-$}}
\newcommand{\hmm}{\mbox{$\et\rightarrow \mu^+\mu^-$}}
\newcommand{\hppee}{\mbox{$\et\rightarrow\pi^+\pi^-e^+e^-$}}
\newcommand{\hppg}{\mbox{$\et\rightarrow \pi^+\pi^-\gamma$}}
\newcommand{\hgg}{\mbox{$\et\rightarrow \gamma\gamma$}}
\newcommand{\hppp}{\mbox{$\et\rightarrow \pi^+\pi^-\pi^0$}}
\newcommand{\hpppo}{\mbox{$\et\rightarrow \pi^0\pi^0\pi^0$}}
\newcommand{\hpppod}{\mbox{$\et\rightarrow \pi^0\pi^0\pi^0_D$}}

\newcommand{\peeg}{\mbox{$\pi^0\rightarrow e^+e^-\gamma$}}
\newcommand{\pee}{\mbox{$\pi^0\rightarrow e^+e^-$}}
\newcommand{\eeg}{\mbox{$e^+e^-\gamma$}}
\newcommand{\et}{\mbox{$\eta$}}
\newcommand{\pdHeh}{\mbox{$pd\to^3$He\,$\eta$}}

\newcommand{\mum}{\mbox{$\mu^-$}}
\newcommand{\mup}{\mbox{$\mu^+$}}

\newcommand{\E}[1]{\mbox{$\times$10$^{#1}$}}

\newcommand{\xpicsize}{\columnwidth}
\newcommand{\IMeeg}{$M(\eeg)$}

\begin{document}
\title{Measurement of $\eta$ meson decays into lepton-antilepton
pairs}

\author{M.~Ber{\l}owski}
\affiliation{The Andrzej Soltan Institute for Nuclear Studies, Warsaw and Lodz, Poland}
\author{Chr.~Bargholtz}
\affiliation{Stockholm University, Stockholm, Sweden}
\author{M.~Bashkanov}
\affiliation{Physikalisches Institut der Universit\"at T\"ubingen, T\"ubingen, Germany}
\author{D.~Bogoslawsky}
\affiliation{Joint Institute for Nuclear Research, Dubna, Russia}
\author{A.~Bondar}
\affiliation{Budker Institute of Nuclear Physics, Novosibirsk, Russia}
\author{H.~Cal\'en}
\affiliation{The Svedberg Laboratory, Uppsala, Sweden}
\author{F.~Cappellaro}
\affiliation{Uppsala University, Uppsala, Sweden}
\author{H.~Clement}
\affiliation{Physikalisches Institut der Universit\"at T\"ubingen, T\"ubingen, Germany}
\author{L.~Demir\"ors}
\affiliation{Institut f\"ur Experimentalphysik der Universit\"at Hamburg, Hamburg, Germany}
\author{C.~Ekstr\"om}
\affiliation{The Svedberg Laboratory, Uppsala, Sweden}
\author{K.~Fransson}
\affiliation{The Svedberg Laboratory, Uppsala, Sweden}
\author{L.~Ger\'en}
\affiliation{Stockholm University, Stockholm, Sweden}
\author{L.~Gustafsson}
\affiliation{Uppsala University, Uppsala, Sweden}
\author{B.~H\"oistad}
\affiliation{Uppsala University, Uppsala, Sweden}
\author{G.~Ivanov}
\affiliation{Joint Institute for Nuclear Research, Dubna, Russia}
\author{M.~Jacewicz}
\affiliation{Uppsala University, Uppsala, Sweden}
\author{E.~Jiganov}
\affiliation{Joint Institute for Nuclear Research, Dubna, Russia}
\author{T.~Johansson}
\affiliation{Uppsala University, Uppsala, Sweden}
\author{S.~Keleta}
\affiliation{Uppsala University, Uppsala, Sweden}
\author{I.~Koch}
\affiliation{Uppsala University, Uppsala, Sweden}
\author{S.~Kullander}
\affiliation{Uppsala University, Uppsala, Sweden}
\author{A.~Kup\'s\'c}
\affiliation{The Svedberg Laboratory, Uppsala, Sweden}
\author{A.~Kuzmin}
\affiliation{Budker Institute of Nuclear Physics, Novosibirsk, Russia}
\author{A.~Kuznetsov}
\affiliation{Joint Institute for Nuclear Research, Dubna, Russia}
\author{I.V.~Laukhin}
\affiliation{Moscow Engineering Physics Institute, Moscow, Russia}
\author{K.~Lindberg}
\affiliation{Stockholm University, Stockholm, Sweden}
\author{P.~Marciniewski}
\affiliation{The Svedberg Laboratory, Uppsala, Sweden}
\author{R.~Meier}
\affiliation{Physikalisches Institut der Universit\"at T\"ubingen, T\"ubingen, Germany}
\author{B.~Morosov}
\affiliation{Joint Institute for Nuclear Research, Dubna, Russia}
\author{W.~Oelert}
\affiliation{Institut f\"ur Kernphysik, Forschungszentrum J\"ulich, J\"ulich, Germany}
\author{C.~Pauly}
\affiliation{Institut f\"ur Experimentalphysik der Universit\"at Hamburg, Hamburg, Germany}
\author{H.~Pettersson}
\affiliation{Uppsala University, Uppsala, Sweden}
\author{Y.~Petukhov}
\affiliation{Joint Institute for Nuclear Research, Dubna, Russia}
\author{A.~Povtorejko}
\affiliation{Joint Institute for Nuclear Research, Dubna, Russia}
\author{R.J.M.Y.~Ruber}
\affiliation{The Svedberg Laboratory, Uppsala, Sweden}
\author{K.~Sch\"onning}
\affiliation{Uppsala University, Uppsala, Sweden}
\author{W.~Scobel}
\affiliation{Institut f\"ur Experimentalphysik der Universit\"at Hamburg, Hamburg, Germany}
\author{R.~Shafigullin}
\affiliation{Moscow Engineering Physics Institute, Moscow, Russia}
\author{B.~Shwartz}
\affiliation{Budker Institute of Nuclear Physics, Novosibirsk, Russia}
\author{T.~Skorodko}
\affiliation{Physikalisches Institut der Universit\"at T\"ubingen, T\"ubingen, Germany}
\author{V.~Sopov}
\affiliation{Institute of Theoretical and Experimental Physics, Moscow, Russia}
\author{J.~Stepaniak}
\affiliation{The Andrzej Soltan Institute for Nuclear Studies, Warsaw and Lodz, Poland}
\author{P.-E.~Tegn\'er}
\affiliation{Stockholm University, Stockholm, Sweden}
\author{P.~Th\"orngren Engblom}
\affiliation{Uppsala University, Uppsala, Sweden}
\author{V.~Tikhomirov}
\affiliation{Joint Institute for Nuclear Research, Dubna, Russia}
\author{A.~Turowiecki}
\affiliation{Institute of Experimental Physics, Warsaw, Poland}
\author{G.J.~Wagner}
\affiliation{Physikalisches Institut der Universit\"at T\"ubingen, T\"ubingen, Germany}
\author{M.~Wolke}
\affiliation{Institut f\"ur Kernphysik, Forschungszentrum J\"ulich, J\"ulich, Germany}
\author{A.~Yamamoto}
\affiliation{High Energy Accelerator Research Organization, Tsukuba, Japan}
\author{J.~Zabierowski}
\affiliation{The Andrzej Soltan Institute for Nuclear Studies, Warsaw and Lodz, Poland}
\author{I.~Zartova}
\affiliation{Stockholm University, Stockholm, Sweden}
\author{J.~Z{\l}oma\'nczuk}
\affiliation{Uppsala University, Uppsala, Sweden}
\collaboration{CELSIUS/WASA Collaboration}

\date{\today}

\begin{abstract}
A  search for rare  lepton decays  of the  $\eta$ meson  was performed
using the WASA detector at  CELSIUS.  Two candidates for double Dalitz
decay $\eta\to e^+e^-e^+e^-$ events  are reported with a background of
1.3$\pm$0.2  events.   This  allows  to  set an  upper  limit  to  the
branching  ratio  of 9.7$\times$10$^{-5}$  (90\%  CL).  The  branching
ratio  for   the  decay   $\eta\to  e^+e^-\gamma$  is   determined  to
$(7.8\pm0.5_{stat}\pm0.8_{syst})\times10^{-3}$ in agreement with world
average value.  An  upper limit (90\% CL) for  the branching ratio for
the $\eta\to e^+e^-$  decay is $2.7\times10^{-5}$ and a  limit for the
sum    of   the    $\eta\to    \mu^+\mu^-\mu^+\mu^-$   and    $\eta\to
\pi^+\pi^-\mu^+\mu^-$ decays is $3.6\times10^{-4}$.
\end{abstract}


\maketitle

\section{Introduction}   \label{introd}

\begin{figure}
\hspace{-0.5cm}a)\includegraphics[width=0.19\textwidth]{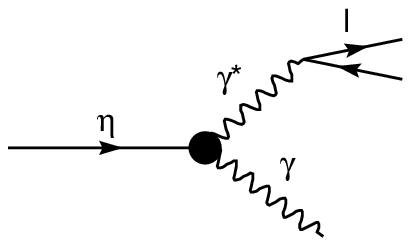}
b)\includegraphics[width=0.19\textwidth]{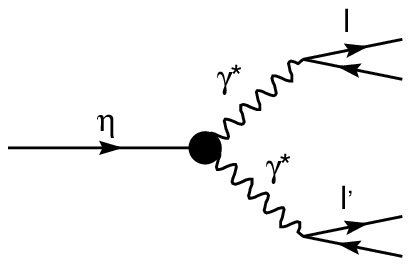}
\hspace*{0.5cm}c)\includegraphics[width=0.2\textwidth]{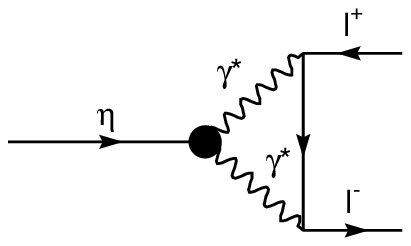}
\caption{\label{fig:hee2b}  Diagrams for a)  single, b)  double Dalitz
decays of a neutral pseudoscalar meson ($\pi^0$, $\eta$ or $\eta'$) and c)
dominating  conventional mechanism  for  decay into  a lepton-antilepton
pair.}
\end{figure}

The  $\eta$  decays with  lepton  pairs  are  closely related  to  the
channels  with   real  photons.   A  direct   consequence  of  Quantum
Electrodynamics  is  that a  process  with  a  real photon  should  be
accompanied by  a process where  a virtual photon  converts internally
into a lepton-antilepton pair~(fig.~\ref{fig:hee2b}a,b).  This fact was
first pointed out by  Dalitz in 1951~\cite{Dalitz:1951aj}.  The decays
can be related  to the corresponding radiative decays  with one or two
photons using Quantum Electrodynamics and by introducing a function of
the   four  momentum   transfer   squared  of   the  virtual   photons
($q_{1,2}^2$): $F(q_1^2,q_2^2,m_\eta^2)$  -- the  transion   {\em Form
Factor} (FF)  (an overview is given  e.g. in \cite{Landsberg:1986fd}).
The $q_{1,2}^2$  for the Dalitz decay  is equal to  the invariant mass
squared of the lepton-antilepton  pair and $q_{1,2}^2\ge 4m_l^2$.  The
FF describes  the structure  of the transition  region and it  is also
used  for the  process  $\gamma^*\gamma^*\to\eta$ where  $q_{1,2}^2<0$
(space-like virtual photons).

\begin{table*}[htpb]
   \begin{tabular*}{\textwidth}{@{\extracolsep{\fill}}lrrl}
    \hline\hline
     Decay mode & BR exp.    & BR theor.&Remarks\\
    \hline
    \heeg    &  $(6.0\pm0.8)\times10^{-3}$  &  $(6.37-6.57)\times10^{-3}$& \\
    \hmmg    &  $(3.1\pm0.4)\times10^{-4}$  &  $(2.10-3.05)\times10^{-4}$&  \\
    \heeee   &  $<6.9\times10^{-5}$         &  $(2.52-2.64)\times10^{-5}$&
Data CMD-2\cite{Akhmetshin:2000bw}\\
    \heemm   &  --     &  $(1.57-2.21)\times10^{-7}$& \\
    \hmmmm   &  --       &  $2.4\times10^{-9}$&  \\
    \hee     &  $<7.7\times10^{-5}$         & $\ge1.7\times10^{-9}$&
Data CLEO II\cite{Browder:1997eu}, Unitarity bound \cite{Bergstrom:1982zq}  \\
    \hmm     &  $(5.8\pm0.8)\times10^{-6}$  &  $\ge 4.3\times10^{-6}$&
Unitarity bound \cite{Bergstrom:1982zq}\\
    \hppee   &  $(4.3\pm1.3\pm0.4)\times10^{-4}$  &  $(3.0-3.6)\times10^{-4}$&
 Data CELSIUS/WASA
   \cite{Bargholtz:2006gz} \\
    \hppmm   &  --       &  $7.5\times10^{-9}$&  \\
    $\eta\to\mu^\pm e^\mp$   &   $<6\times10^{-6}$      &  0&Violates Lepton Flavor  \\
    \hline\hline
   \end{tabular*}
   \caption{\label{tab:0}The measured and calculated branching ratios for 
   different
   \et~ decay channels with lepton-antilepton pair(s). The data are from
   \cite{Yao:2006px} if not stated otherwise.  The upper limits
   are for 90\% CL. Calculations for single and double Dalitz decays are
   from \cite{Jarlskog:1967aa, Picciotto:1993aa,  Faessler:1999de,
   Bijnens:1999jp, Borasoy:2007dw}. }
\end{table*}
Experimental information  is scarce  even for not  so rare  \et~ meson
decays  with electron-positron pair(s)  as seen  in table~\ref{tab:0},
where  measured and  predicted branching  ratios (BR)  are summarized.
Even the branching  ratio for the \heeg~ decay is  known with a rather
large uncertainty  $(6.0\pm0.8)\times10^{-3}$ \cite{Yao:2006px}. It is
worth noting that the quoted value  was obtained as the average of two
experimental  results with a rather large scale factor --  $1.4$.  The
recent result  from the CLEO  Collaboration $(9.4\pm0.7)\times10^{-3}$
\cite{Lopez:2007pp} is  larger by three standard  deviations.  None of
the \et~decays  with double  lepton-antilepton pairs were  observed so
far.  The  decays were studied  theoretically already 40 years  ago by
Jarlskog  and Pilkhun  \cite{Jarlskog:1967aa} assuming  a FF  equal to
one.  The effect of the FF on  the BR is expected to be less than 10\%
for  the   decay  \heeee\  \cite{Bijnens:1999jp}.    For  decays  with
$\mu^+\mu^-$ pair(s)  the influence is  larger since only  large $q^2$
values are probed.

Decays of  neutral pseudoscalar mesons into  a lepton-antilepton pair,
$P\rightarrow \ell^+\ell^-$, represent a potentially important channel
to   look  for   effects  of   physics  beyond   the   Standard  Model
\cite{Bergstrom:1982zq}.  The  dominant mechanism within  the Standard
Model  is   a  second  order   electromagnetic  process,  additionally
suppressed  by helicity  conservation, involving  two  virtual photons
$P\rightarrow\gamma^*\gamma^*$ shown in fig.~\ref{fig:hee2b}c.  Due to
the loop appearing in the diagram the decay is sensitive to the values
of   the   FF   for   any   $q_{1,2}^2$  of   the   photons   in   the
loop~\cite{Drell:1959aa}.  The  imaginary part of  the decay amplitude
can be  uniquely related to the  decay width of the  \hgg\ decay.  The
experimental  value of  $\Gamma(\hgg)$  leads to  a  lower limit  (the
unitarity       bound)       of       the       branching       ratio:
BR$(\hee)\geq1.7\times10^{-9}$  when  the   real  part  of  the  decay
amplitude  is   neglected  \cite{Bergstrom:1982zq,  Landsberg:1986fd}.
This value is  much lower than for other decays of  \po\ and \et\ into
lepton-antilepton pairs.  This makes the \hee~ decay rate sensitive to
a possible exotic contribution.  The best experimental upper limit for
the    $BR(\hee)$    comes   from    the    CLEO   II    collaboration
\cite{Browder:1997eu}   and  is  four   orders  of   magnitude  higher
(table~\ref{tab:0}).   The  decays  \pee,  \hmm~ and  \hee~  are  also
important in  order to estimate  long range contribution to  the decay
$K_L\to\mup\mum$.  The loop diagram of the short-distance amplitude is
sensitive to the presence of a  virtual top quark and could be used to
improve  the knowledge  on the  $|V_{td}|$ element  of the  CKM matrix
\cite{GomezDumm:1998gw,Isidori:2003ts}.

The real  part of the  amplitude of the  \hee~ decay can  be estimated
using   the  measured   value   of  BR(\hmm)   \cite{GomezDumm:1998gw,
Savage:1992ac, Ametller:1993we}. The assumption that the ratio between
Im and Re parts of the amplitudes  for the decays is the same leads to
the  prediction $BR(\hee)\approx  (6\pm 0.2)\times  10^{-9}$.   A new,
unknown process  could increase the  value.  Recently the  interest in
the  decays  was  revived due  to  the  observed  excess rate  of  the
$\pi^0\to  e^+e^-$ decay  \cite{Abouzaid:2006kk} with  respect  to the
Standard  Model   predictions  \cite{Dorokhov:2007bd}  what  triggered
theoretical speculations that the excess  might be caused by a neutral
vector  meson responsible for  annihilation of  a neutral  scalar dark
matter  particle \cite{Kahn:2007ru}.  The  consequence could  be large
(even  an oder of  magnitude) enhancement  of the  $\eta \to  e^+ e^-$
decay rate.

The plan of this paper is the following: In part II, the experiment is
described and  the data selection  is presented.  In section  II.A the
\hpppo\  decay  where one  of  the  neutral  pions decays  via  \peeg\
(\hpppod)  is   presented.   The  process   is  used  to   verify  the
understanding of the detector response for electrons and positrons and
to provide normalization for the  BR of leptonic \et\ decays.  This is
an extension  of the systematical studies from  a previous publication
that  used the same  data sample~\cite{Bargholtz:2006gz}.   In section
III.A the Dalitz decay \heeg~  is considered and the BR is determined.
In sections III.B  to III.D the results of the  search for the \heeee,
\hmmmm, \hppmm\ and \hee\ decays are presented.

\section{The Experiment \label{exp}}
\begin{figure*}[htpb]
\includegraphics[width=0.9\textwidth]{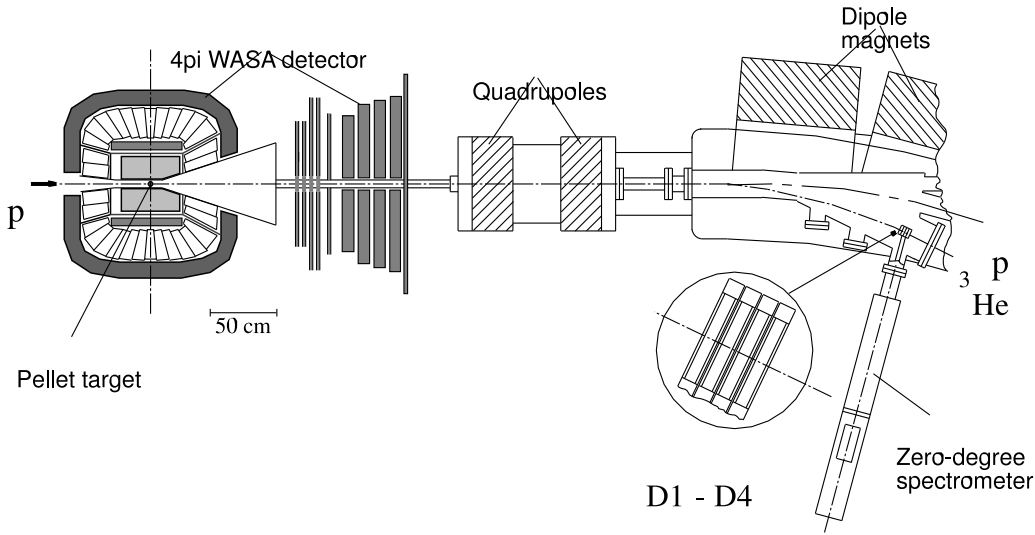}
\caption{ \label{fig:ts_bw2}
The WASA detector with zero-degree spectrometer using CELSIUS dipoles.}
\end{figure*}
The experiment was  performed at the CELSIUS storage  ring in Uppsala,
using     the    WASA    detector     setup    (fig.~\ref{fig:ts_bw2})
\cite{Zabierowski:2002ah}.  Protons  with a kinetic energy  of 893 MeV
interacted  with frozen  droplets of  deuterium \cite{Ekstrom:2002ai}.
The \et~  mesons were  produced in the  reaction \pdHeh\ close  to the
\et~  production  threshold.   The  detection  of  $^3$He  ions  in  a
zero-degree   spectrometer  (tagging   detector)   provided  a   clean
\et~trigger independent of decay channel \cite{Bargholtz:1997zr}.  The
$^3$He  ions  were identified  and  their  energy  was measured  which
allowed a  clean selection of  the \pdHeh\ reaction with  a background
(mainly  due  to  $pd\to^3$He$\pi\pi$  reaction) of  about  1\%.   The
tagging detector provided  a few triggers per second  (at a luminosity
of $ 5\times 10^{30}cm^{-2}s^{-1}$),  yielding on average one recorded
\et~  event   per  second.   During   the  two  weeks   of  experiment
(distributed over a period of  half a year) nearly $3\times 10^5$ \et~
events were collected.

The charged \et~ decay products  were tracked using a cylindrical mini
drift chamber (MDC), consisting of 17 layers of thin-walled (25$\mu$m)
aluminized mylar tubes and built around a beryllium beam pipe of 60 mm
diameter  with wall  thickness  of only  1.2  mm (3.4\E{-3}  radiation
lengths).   Since  the target  deuterium  droplets  have  a radius  of
17$\mu$m  (2\E{-6}  radiation  lengths)  the  beam pipe  is  the  most
important  source of  photon conversion  background.  For  example the
fraction of \ee\ pairs from $\eta\to\gamma\gamma$ with external photon
conversion in the  beam pipe to the Dalitz pairs  from \heeg\ is about
60\%.   This background could  be further  suppressed by  checking the
reconstructed position of the vertex of a pair.  The above features of
the WASA detector are crucial  for the investigation of reactions with
\ee\ pairs.   The MDC is  placed inside of a  superconducting solenoid
which provides  a magnetic field  of 1T.  The  MDC is surrounded  by a
barrel of plastic scintillators used mainly to define event start time
for  drift  time  reconstruction   in  the  MDC.   An  electromagnetic
calorimeter consisting of 1012  CsI(Na) crystals measures the energies
of photons and their impact points.

Tracks of electrons and charged pions are reconstructed in MDC with an
efficiency of about 80\% if the transverse particle momenta are larger
than about  20 MeV/c.  One  should stress that  even lepton-antilepton
pairs with parallel  momenta (and thus minimal value  of the invariant
mass)  could   be  efficiently  measured   in  the  MDC.    The  track
reconstruction algorithm for  the MDC used in the  present studies was
based on  a global method of  pattern recognition in  which a constant
magnetic  field   was  assumed.    The  position  resolution   of  the
reconstructed vertex is about 0.05  cm FWHM in the plane perpendicular
to the beam and 0.7 cm FWHM along the beam.

In the  offline analysis,  events with at  least two  charged particle
tracks reconstructed in the MDC  were required. Events with the tracks
originating  far   from  the  beam  target   interaction  region  were
rejected. Hit  clusters in the calorimeter,  without associated tracks
in the MDC and with energy  deposit larger than 20 MeV were assumed to
originate  from  photons.    Only  events  containing  decay  particle
candidates  with balanced  electric charge  were accepted  for further
analysis.   The results  on  the \hppee~  decay  channel were  already
presented  earlier~\cite{Bargholtz:2006gz}.   Events  with a  pair  of
charged  decay   products  with  opposite  electric   charges  can  be
attributed either to the decay channels with two charged leptons or to
more frequent channels with two charged pions (\hppg~ and \hppp).

The following variables are used in the further data analysis:
\begin{itemize}
\item The  invariant mass  of a pair  of oppositely  charged particles
($M_{ee}$).  The electron  mass is used in the  calculations.  A clear
peak at the lowest value is expected for \ee~ pairs from Dalitz decays
and from conversion of real photons in the detector material.
\item  The total invariant  mass of  all reconstructed  decay products
(for example \IMeeg\ or $M(3\po)$).   It was required to be consistent
with the \et~ mass.
\item The missing mass of all decay products ($MM_{\et}$).  It should,
within  errors, be  equal to  the mass  of the  $^3$He  nucleus (2.808
GeV/c$^2$).
\item  The ratio  between the  momentum measured  in the  MDC  and the
energy  of the  shower in  the electromagnetic  calorimeter associated
with the charged track ($R_{p/E}$).  It permits to distinguish between
$e^\pm$  and $\pi^\pm$  when the  particles reach  the electromagnetic
calorimeter. 
\item  The opening  angle between  two reconstructed  real  or virtual
photons  ($\theta_{\gamma\gamma*}$  or $\theta_{\gamma*\gamma*}$)  and
the     relative    azimuthal     angle     between    the     photons
($\Delta\phi_{\gamma\gamma*}$  or  $\Delta\phi_{\gamma*\gamma*}$). The
angles are given in the laboratory frame.
\end{itemize}

The  separation of  electrons  from pions  relies  in the  end on  the
kinematics of the reactions studied. Due to the large mass difference,
the  energy momentum  conservation  is violated  with  the wrong  mass
assignment.  For  example was the contribution of  the background from
$pd\to^3$He$\pi^+\pi^-$ to  the final selection of  the \hee\ reaction
found to be negligible.  Conversely do neither the kinematics or other
particle  identification methods  allow us  to distinguish  pions from
muons in the studied channels.

\subsection{Normalization: \hpppod\ decay}

In order  to normalize the branching  ratios of the  \et~ meson decays
involving an  \ee\ pair, a monitoring  process is needed  to check the
reconstruction  efficiency  for  electrons  and  positrons.   This  is
specially  important since the  experiment was  split into  short time
slices distributed over a longer  time.  This data sample was analyzed
already for  a previous  paper \cite{Bargholtz:2006gz} on  the \hppee\
decay  mode  where   the  \hppp\  decay  was  used   for  the  primary
normalization.  A cross check was done using \eeg\ decays assuming the
Particle  Data Group (PDG)  \cite{Yao:2006px} value  for this  BR.  To
reduce  the  systematical uncertainty  and  in  order  to be  able  to
determine the BR(\eeg) we used  in this paper both \hppp\ and \hpppod\
decays for the normalization.

\begin{figure}[htbp]
\includegraphics[width=\xpicsize]{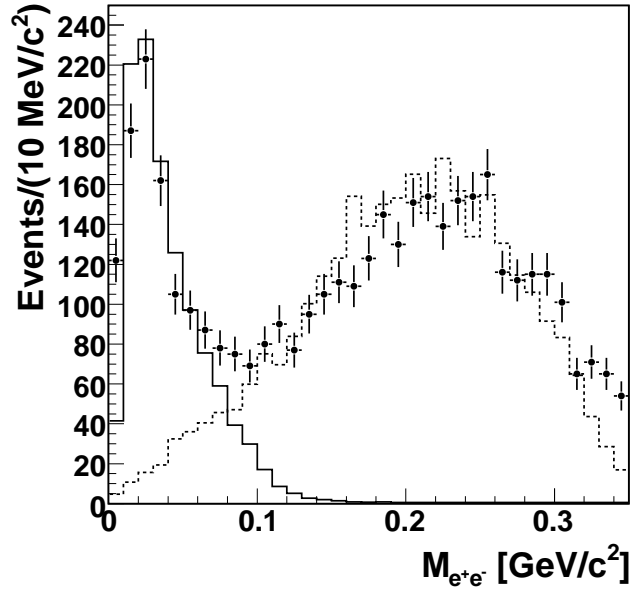}
\caption{ \label{fig:mee_3pi0}  Invariant mass of  the \ee~ candidates
for events with  more than three neutral hit clusters.  Points -- the
experimental data, solid line -- MC simulation for \hpppod\
decay,   dotted line   --  background   from   decays  involving
a misidentified $\pi^+\pi^-$ pair.}
\end{figure}

Dalitz decays  of at  least one  of the three  \po's from  the \hpppo\
decays provide an abundant data set of events with five photons and an
electron-positron pair -- \hpppod.  The Dalitz decay of the \po\ meson
has been studied in  detail both theoretically and experimentally.  To
select a data sample of \hpppod\ events we required:
\begin{itemize}
\item at least two tracks from particles with opposite charges
\item more than three neutral hit clusters in the calorimeter.
\end{itemize}
In  fig.~\ref{fig:mee_3pi0} the  experimental $M_{ee}$ distribution for
such events  is plotted.
The peak  at low masses is  attributed to the \ee~  pairs from \hpppo\
decays  with internal  or external  conversion of  one of  the photons
(solid line  in the  fig.~\ref{fig:mee_3pi0}).  The maximum  at larger
masses  is due to  \et~ decays  with a  $\pi^+\pi^-$ pair,  mainly the
\hppp\  decay.  The relative  normalization of  the decays  differs by
15\% from what is expected from the branching ratios.  This difference
is attributed to the lower reconstruction efficiency for electrons and
positrons than for charged pions in the MDC.

\begin{figure}[htbp]
\includegraphics[width=\xpicsize]{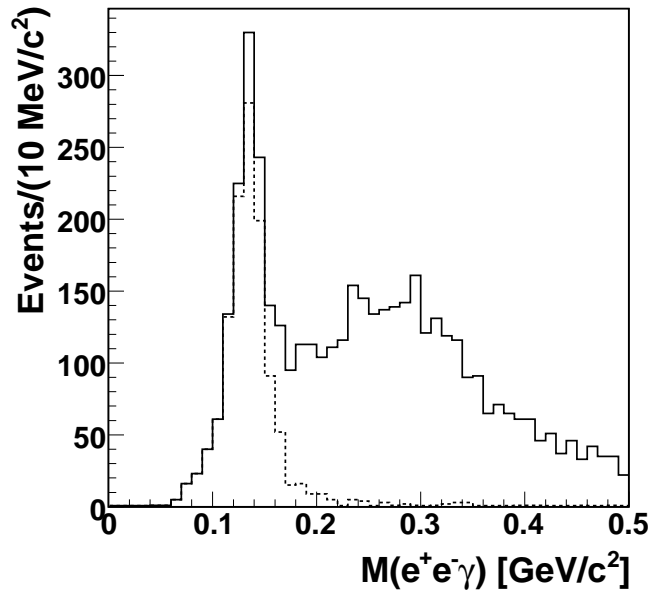}
\caption{  \label{fig:3pi0_meeg_exp}  The  \eeg~  invariant  mass  for
events  with at  least four  neutral hit  clusters. Full  line  -- all
events,  dotted line  -- events  with  \ee\ invariant  mass less  than
$0.1~GeV$. The photon  giving the \IMeeg\ mass closest  to the neutral
pion mass was selected.}
\end{figure}

\begin{figure}[htbp]
\includegraphics[width=\xpicsize]{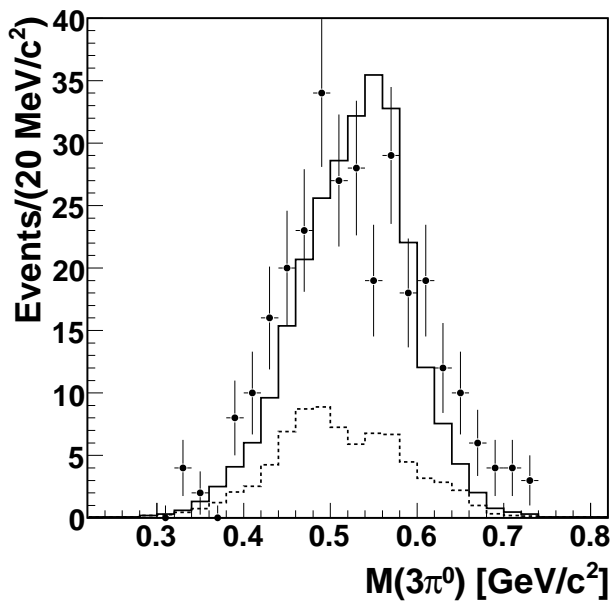}
\caption{ \label{fig:3pi0_full} The  $M(3\po)$ distribution for events
with three neutral pions reconstructed  after all cuts.  Points -- the
experimental  data, solid line  -- MC  simulation for  the sum  of the
\hpppod\  decay  and  the  background,  dashed line  --  \hpppo~  with
external conversion.}
\end{figure}

The identification of  the \hpppod\ decay channel is  confirmed by the
reconstruction  of the  invariant mass  of the  \eeg\  system (\IMeeg)
where the photon leading to the  mass value closest to the \po~mass is
selected (fig.~\ref{fig:3pi0_meeg_exp}).   The \IMeeg\ distribution is
peaked at  the \po~mass  when the $M_{ee}<0.1$~GeV/c$^2$  condition is
applied.   Fig.~\ref{fig:3pi0_full} shows  the invariant  mass  of the
three \po's for the events  where all pion energies are below 0.2~GeV.
For the final data sample it was required that the missing mass of the
system  of  all  decay  products  is in  the  range  2.5~GeV/c$^2$  to
3.0~GeV/c$^2$ and the reconstructed  emission angle of the $\et$ meson
is less than  60\degr.  Assuming that all remaining  events are due to
the  decays of  \et~ into  three neutral  pions, the  total  number of
$\eta$ mesons $N_{\et}$ is calculated from the formula:
\begin{equation}
N_{\et}=\frac{N_D}{(1-(1-p)^3){\cal A}\ BR(\hpppo)}
\end{equation}
where    $N_D$    is   the    number    of    the   observed    events
(fig.~\ref{fig:3pi0_full}  after background subtraction)  and $p\equiv
BR(\peeg)=(1.198\pm0.032)\%$  \cite{Yao:2006px}.  The  product  of the
detector acceptance and the reconstruction efficiency (${\cal A}=(13.8
\pm 2.0)\%$)  was extracted from  a MC simulation assuming  the Vector
Meson  Dominance model  Form Factor  for  the $\pi^0$.   The value  of
$BR(\hpppo)$     is    precisely    known     --    $(32.51\pm0.28)\%$
\cite{Yao:2006px}.
The extracted $N_{\et}=256000\pm  18000$ agrees with the value
from  our previous paper  \cite{Bargholtz:2006gz} within  two standard
deviations. Then, for  the $BR$ normalization in the  present paper we
use the weighted mean value: $N_{\et}=241000\pm 13000$.

\section{Results}
\subsection{Single Dalitz decay \heeg}

\begin{figure}[htbp]
\includegraphics[width=\xpicsize]{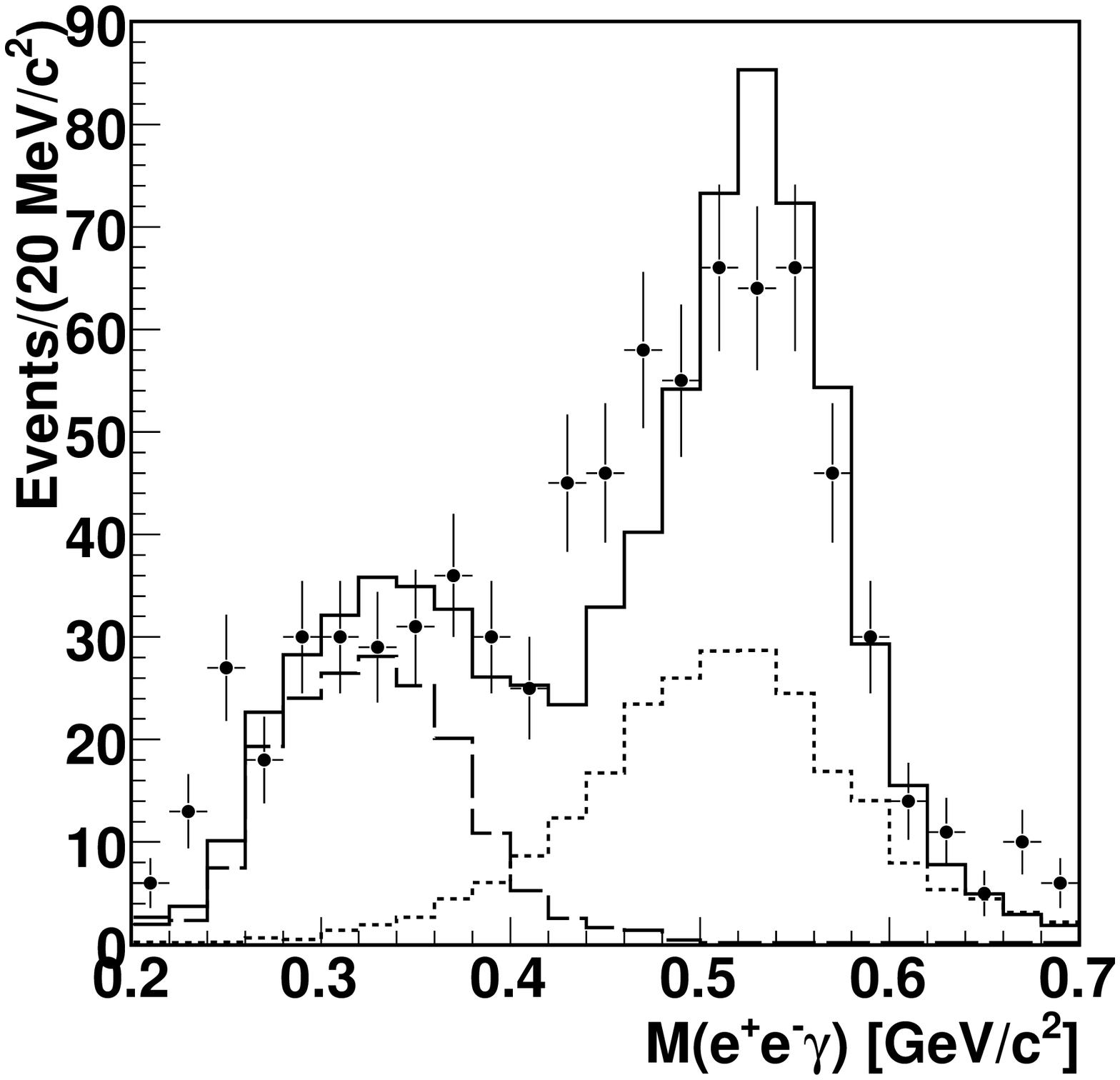}
\caption{  \label{fig:meeg_iden} The  \IMeeg\ distribution  for events
with $M_{ee}<0.125$~GeV/c$^2$ after particle identification. Points --
data, solid  line is the sum  of MC simulations of  the signal (\heeg)
and the background. Dashed line -- contribution from \hppp\ and \hppg\
decays; dotted  line -- \hgg\ with  one of the  photons converted into
\ee\ pair in the detector material. }
\end{figure}

Fig.~\ref{fig:meeg_iden}  shows  the  invariant mass  distribution  of
\eeg\  candidates (\IMeeg) selected by the following conditions:
\begin{itemize}
\item at least two tracks from particles with opposite charges
\item  for  tracks  with  matched  hit  clusters  in  the  calorimeter
the condition $R_{p/E}<1.65$ was applied
\item  $M_{ee}<0.125$~GeV/c$^2$
\item a neutral hit cluster with energy  deposit larger than 180~MeV.
\end{itemize}
A clear  signal at the \IMeeg~  around the $\et$~ meson  mass is seen.
The solid line in  Fig.~\ref{fig:meeg_iden} represent MC simulation of
signal and a sum of all background contributions.  External conversion
of one  of the photons from  \hgg\ decay comprises  the most important
background   as  discussed   in   the  section   \ref{exp}.   The   MC
underestimates the detector resolution in the \IMeeg.  The discrepancy
is  caused by  not optimal  calibration  of the  calorimeter which  is
difficult  to improve  since the  data taking  was distributed  over a
longer  time.  We  have checked  the influence  of the  effect  on the
extracted  values   of  the  BR   by  artificially  smearing   the  MC
distributions to match the experimental data.

\begin{figure}[htbp]
\includegraphics[width=0.5\textwidth]{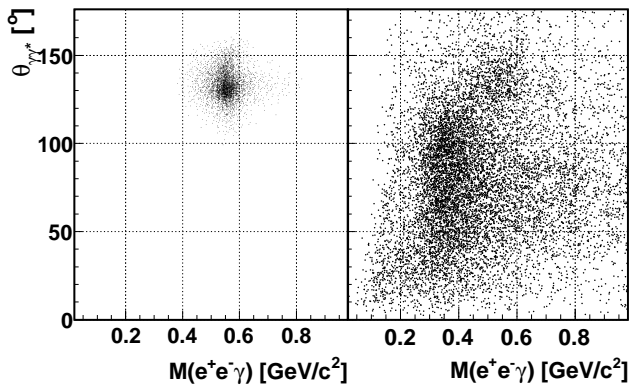}
\caption{ \label{fig:imeeg_degeeg}  $\theta_{\gamma\gamma*}$ vs \IMeeg~
before  cleaning cuts and without
particle identification:  left  --  MC simulation  for  \heeg, right  --
experimental data candidates.}
\end{figure}

\begin{figure}[htbp]
\includegraphics[width=\xpicsize]{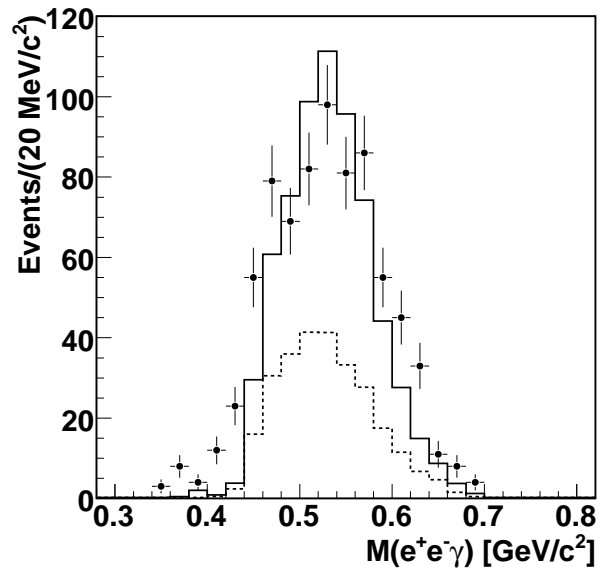}
\caption{  \label{fig:eeg_final} The  \IMeeg\  distribution after  the
final  selection.  Points  --  experimental  data, solid  line  --  MC
simulation of \heeg, dotted line -- MC simulation of \hgg\ with photon
conversion in the detector material.}
\end{figure}

There  are no restrictions  on the  number of  low energy  neutral hit
clusters  since   due  to  electron  or  photon   interaction  in  the
calorimeter an additional hit cluster  can be created.  For about 22\%
of   the  events,   an  additional   low-energy  neutral   cluster  is
reconstructed.  The photon candidate for the \heeg\ decay was selected
by  the  requirement that  the  $\Delta\phi_{\gamma\gamma*}$ angle  is
closest  to  180\degr.   The  signature  of the  \heeg\  decay  is  an
energetic   photon   ($E_{\gamma}>0.18$~GeV).    The   opening   angle
$\theta_{\gamma\gamma*}$   is   distributed   between  110\degr\   and
150\degr\ peaking around 130\degr.  Fig.  \ref{fig:imeeg_degeeg} shows
$\theta_{\gamma\gamma*}$ versus \IMeeg.
A constraint on the angle: 100\degr$<\theta_{\gamma\gamma*}<$160\degr\
together with  a condition on the  overall missing mass  for the decay
system  2.65~GeV/c$^2<MM_{\et}<$2.90~GeV/c$^2$ cleans the  data sample
significantly. This allows to  release the condition on $R_{p/E}<1.65$
and this  increases the acceptance  since not all $e^+e-$  from \heeg\
decay reach the calorimeter.   Finally 729 events with \IMeeg\ between
0.40  and 0.64~GeV/c$^2$  are identified.   The total  contribution of
background  (mainly from  \hppg,  \hppp~  and \hgg~  with  one of  the
photons  converting  into  \ee\  pair  in the  detector  material)  is
estimated to  $294\pm15$ events.  Fig.   \ref{fig:eeg_final} shows the
\IMeeg\  distribution  after  applying  all selection  cuts  mentioned
above.

\subsection{The decay \heeee}

\begin{figure}[htbp]
\includegraphics[angle=270,clip=,width=\xpicsize]{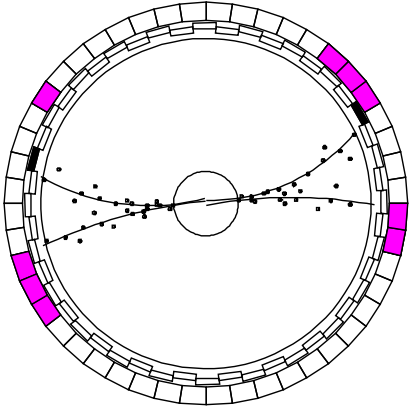}
\caption{\label{fig:event}(Color online) Event  display for an \heeee~
candidate event.  The shaded area in the outermost ring represents the
projection of the  hit calorimeter crystals (the size  of the crystals
and the  radial position  of the  front faces are  not to  scale). The
lines represent the reconstructed  tracks from the pattern recognition
program. In  addition to  layers with straws  along the beam,  the MDC
includes  twisted  layers  which   cause  the  spread  of  points  for
forward/backward going tracks. }
\end{figure}

In a search for the  \heeee\ decay, events with exactly two positively
and two negatively  charged particle tracks in the  MDC were selected.
According to the simulations  11\% of the reconstructed \heeee\ events
should fulfill the following criteria:
\begin{itemize}
\item the relative  angle between electron and positron  in both pairs
is smaller than 40\degr
\item the opening angle between the  momenta of the two \ee\ pairs is
in the interval 110\degr\ to 170\degr
\item
the \et~ meson emission angle is smaller than 45\degr
\item the missing transverse momentum is less than 0.3 GeV/c.
\end{itemize}
In  the data only  two events  passed all  selection cuts.   The event
display   for    one   of   the    two   candidates   is    shown   in
fig.~\ref{fig:event}.

The  background  is estimated  to  $1.3\pm0.2$  events and  originates
mainly from single Dalitz decay \heeg\ with the photon converting into
an $e^+e^-$ pair in the detector material.

\subsection{The decays \hmmmm\ and \hppmm}
The decays \hmmmm\  and \hppmm\ have very similar  kinematics.  In the
analysis we have  focused on the \hmmmm\ decay but  it is not possible
to distinguish  the two  in the present  analysis.  One starts  with a
similar  sample of  events as  for  the \heeee\  decay analysis:  four
tracks from  charged particles with  charge balance.  The  events with
neutral hit clusters  of energy larger than 20~MeV or  with a track in
the Forward  Detector (detection angle  2\degr--17\degr) are rejected.
The  kinematics is  checked assuming  muon mass  for the  four charged
particles.   The opening  angle between  the momenta  of the  two muon
pairs  is  required  to  be  in the  interval  26\degr--163\degr.   No
candidate  event for  the discussed  decay channels  is left  for four
candidate muons  with invariant mass  less than 0.625~GeV/c$^2$  and a
missing mass MM$_{\et}$ greater than 2.32~GeV/c$^2$.
\subsection{The decay \hee}

Events with two  tracks from charged particles of  opposite charge are
considered.   The  \hee~ decay  has  a  distinctive  signature in  the
\pdHeh\ reaction close to threshold: the emitted electron and positron
have large  energies ($E>150$  MeV), are co-planar  with the  beam and
have a large opening angle (about 130\degr).
\begin{figure}[htbp]
\includegraphics[width=\xpicsize]{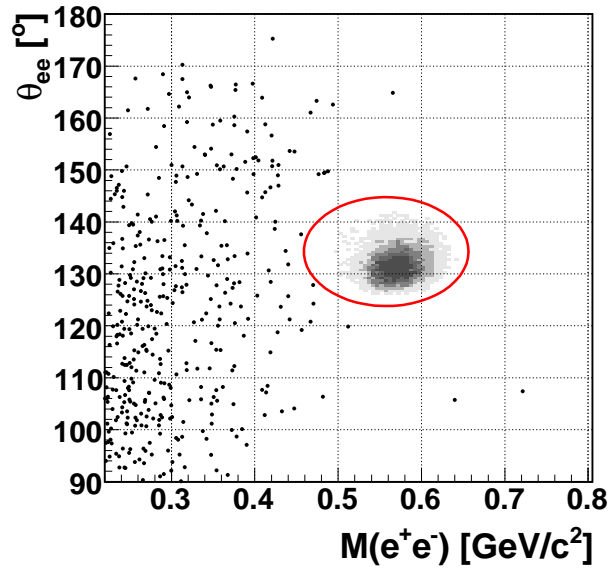}
\caption{   \label{fig:ee2}(Color  online)   Opening   angle  between
electron  and  positron  tracks  vs  $M_{ee}$  for  \hee~event  sample
selection: scatter plot  -- data; shaded area --  MC simulation of the
\hee~decay.  A cut  corresponding to the ellipse shown  in the figure,
selects 71\% of the simulated \hee\ events accepted in the plot.}
\end{figure}
In  fig.~\ref{fig:ee2}, the  \ee~  invariant mass  is  presented as  a
function of  the opening angle  between the electron and  positron for
the  whole  data  set.   The  region of  the  simulated  signal  after
reconstruction cuts is also shown.   There are no events in the region
where the majority of the \hee~ signal is expected: \ee\ opening angle
in   the  interval  120\degr--160\degr,   $M_{ee}>0.49$~GeV/c$^2$  and
fulfilling the particle identification criteria $0.5<R_{p/E}<1.65$.

\subsection{Discussion}
The  results of  the experiment  are summarized  in table~\ref{tab:1}.
The values of the branching ratios are presented in table~\ref{tab:2}.
\begin{table}[htpb]
  \begin{tabular*}{\xpicsize}{@{\extracolsep{\fill}}lrrr}
    \hline\hline
     Decay mode & ${\cal A}$ &  Events      &  Events    \\
                &          & background & observed\\
    \hline
    1. \heeee & (11$\pm$1)\%  & 1.3$\pm$0.2  &  2  \\
    2. \hmmmm & ( 5$\pm$1)\%  & 1.4$\pm$0.9  &  0  \\
    3. \hppmm & ( 5$\pm$1)\%  & 1.4$\pm$0.9  &  0  \\
    4. \hee   & (36$\pm$3)\%  & 0.4$\pm$0.1  &  0  \\
    5. \hppee & (16$\pm$1)\%  & 7.7$\pm$2.0 &  24 \\
    6. \heemm & (16$\pm$2)\%  & 21.0$\pm$2.5&  24 \\
    7. \heeg  & (23$\pm$2)\%  & 294$\pm$15  & 729 \\
    \hline\hline
   \end{tabular*}
   \caption{\label{tab:1}The    detector    acceptance   ${\cal    A}$
   (reconstruction  efficiency  included),   expected  number  of
   background events  and the  number of the  observed events
   after all selection cuts.}
\end{table}
\begin{table}[htpb]
   \begin{tabular*}{\xpicsize}{@{\extracolsep{\fill}}lcc}
    \hline\hline
     Decay mode &BR&BR limit\\
                &&90\% CL\\
    \hline
    1. \heeee &(2.7$^{+2.1}_{-2.7}$$_{stat}\pm0.1_{syst}$)\E{-5}&
    $<$9.7\E{-5}\\
    2. \hppmm &--&$<$3.6\E{-4}\\
    3. \hmmmm &--&$<$3.6\E{-4}\\
    4. \hee   &--&$<$2.7\E{-5}\\
    5. \hppee &$(4.3^{+2.0}_{-1.6}$$_{stat}\pm 0.4_{syst})$\E{-4}&--\\
    6. \heemm &--&$<$1.6\E{-4}\\
    7. \heeg  &(7.8$\pm$0.5$_{stat}\pm$0.8$_{syst}$)\E{-3}&--\\
    \hline\hline
   \end{tabular*}
   \caption{\label{tab:2}  Final results for  the branching  ratios of
lepton \et\ decays.  }
\end{table}
The  confidence limits and  intervals for  decays 1--4  were extracted
using  Feldman  and  Cousins   prescription  for  small  signals  with
background \cite{Feldman:1997qc}.

 The  systematical errors were  obtained by  varying cuts  applied for
selection  of the  channels and  comparison with  Monte  Carlo studies
including e.g.  different  assumptions on photon energy reconstruction
in the  calorimeter and on  the contribution of interaction  with rest
gas.  The main sources of the systematical uncertainty are:
\begin{enumerate}
\item Uncertainty on the total number of the $\eta$ mesons in our data
sample.   That value  is dominated  by a  limited number  of collected
\hpppod~ and  \hppp~ decays. Moreover  the systematical errors  of the
acceptance  and reconstruction  efficiency and  BR for  these channels
were taken into account.  This  contributes with 5.4\% to the relative
uncertainty of the BR for all channels.
\item Uncertainty of the background contribution.
\item Acceptance and reconstruction efficiency uncertainty for a given
channel.   By   using  simultaneously  collected   data  with  similar
topologies for normalization, the contribution of uncertainty of track
reconstruction efficiency is partly canceled.
\end{enumerate}
For  example in the  case of  \heeg~ decay  the systematical  error is
dominated by the uncertainty  in the acceptance ($\Delta{\cal A}/{\cal
A}=8\%$) and it is estimated from the discrepancy between the data and
the MC.

The  extracted  signal   for  the  \heeg\  decay,  $435\pm27_{stat}\pm
15_{syst}$    events,   leads    to    BR(\heeg)=$(7.8   \pm0.5_{stat}
\pm0.8_{syst})\times10^{-3}$.   This is  20\% larger  than theoretical
estimates.  The result is in between the PDG value and the latest CLEO
result \cite{Lopez:2007pp}.

The attempt to extract a branching ratio from the observed two \heeee\
event  candidates  leads to  the  value (2.7$^{+2.1}_  {-2.7}$$_{stat}
\pm0.1_{syst}$)\E{-5}  which  is in  good  agreement with  theoretical
estimates.  However due to non negligible background the value is also
consistent with zero.  If instead  one assumes that the events are due
to background, the upper  limit is $9.7\times10^{-5}$ (90\% CL).  This
improves     slightly     the     previous    limit     from     CMD-2
\cite{Akhmetshin:2000bw}.  The background  is mainly due to conversion
of the  photon from  the \heeg\ decay  in the  beam tube. It  could be
reduced by  checking the  position of the  reconstructed vertex  or by
selecting events with larger invariant masses of the \ee\ pairs.  This
however decreases  the acceptance significantly and could  not be done
in the present study.

The extracted upper limit for BR(\hee) is $2.7\times10^{-5}$ (90\% CL)
and  is  two times  lower  than  the previous  one  from  the CLEO  II
experiment.

We also report on the first  search for the decays \hmmmm~ and \hppmm.
Since the decays can not be distinguished in the present data analysis
an upper  limit of $3.6\times10^{-4}$ (90\%  CL) can be  given for the
sum  of the  decay branching  ratios.  Similarly  \hppee\  and \heemm\
decays  were not  distinguished in  the previous  analysis  of \hppee\
decay  \cite{Bargholtz:2006gz}.  However  the branching  ratio  of the
\heemm\  decay is  expected to  be  three orders  of magnitude  lower.
Assuming  that  BR(\hppee) is  given  as  the  average of  theoretical
predictions,         (3.3$\pm$0.3)\E{-4}        \cite{Jarlskog:1967aa,
Picciotto:1993aa,  Faessler:1999de, Borasoy:2007dw},  and  taking into
account    the    other   sources    of    background   reported    in
\cite{Bargholtz:2006gz} a limit for  the BR(\heemm) to 1.6\E{-4} (90\%
CL) is obtained.  For consistency reasons we have also reevaluated the
result on  BR(\hppee) using the  Feldman and Cousins approach  and the
improved normalization (table~\ref{tab:2}).

\acknowledgments

We are grateful to the  personnel at The Svedberg Laboratory for their
support during  the course of  the experiment.  The  financial support
from the  Knut and Alice  Wallenberg Foundation (Sweden),  the Swedish
Research  Council and  the G\"oran  Gustafsson Foundation  (Sweden) is
acknowledged.   This work has  also been  supported by  BMBF (Germany)
(grants 06HH152,  06TU261), by  Russian Foundation for  Basic Research
(grant  RFBR  02-02-16957)  and  by the  European  Community  Research
Infrastructure Activity under  FP6, Hadron Physics, RII-CT-2004-506078
and HPRI-CT-1999-00098.

\bibliography{eeg}

\begin{thebibliography}{25}
\expandafter\ifx\csname natexlab\endcsname\relax\def\natexlab#1{#1}\fi
\expandafter\ifx\csname bibnamefont\endcsname\relax
  \def\bibnamefont#1{#1}\fi
\expandafter\ifx\csname bibfnamefont\endcsname\relax
  \def\bibfnamefont#1{#1}\fi
\expandafter\ifx\csname citenamefont\endcsname\relax
  \def\citenamefont#1{#1}\fi
\expandafter\ifx\csname url\endcsname\relax
  \def\url#1{\texttt{#1}}\fi
\expandafter\ifx\csname urlprefix\endcsname\relax\def\urlprefix{URL }\fi
\providecommand{\bibinfo}[2]{#2}
\providecommand{\eprint}[2][]{\url{#2}}

\bibitem[{\citenamefont{Dalitz}(1951)}]{Dalitz:1951aj}
\bibinfo{author}{\bibfnamefont{R.~H.} \bibnamefont{Dalitz}},
  \bibinfo{journal}{Proc. Phys. Soc.} \textbf{\bibinfo{volume}{A64}},
  \bibinfo{pages}{667} (\bibinfo{year}{1951}).

\bibitem[{\citenamefont{Landsberg}(1985)}]{Landsberg:1986fd}
\bibinfo{author}{\bibfnamefont{L.~G.} \bibnamefont{Landsberg}},
  \bibinfo{journal}{Phys. Rept.} \textbf{\bibinfo{volume}{128}},
  \bibinfo{pages}{301} (\bibinfo{year}{1985}).

\bibitem[{\citenamefont{Akhmetshin et~al.}(2001)}]{Akhmetshin:2000bw}
\bibinfo{author}{\bibfnamefont{R.~R.} \bibnamefont{Akhmetshin}}
  \bibnamefont{et~al.} (\bibinfo{collaboration}{CMD-2}),
  \bibinfo{journal}{Phys. Lett.} \textbf{\bibinfo{volume}{B501}},
  \bibinfo{pages}{191} (\bibinfo{year}{2001}), \eprint{hep-ex/0012039}.

\bibitem[{\citenamefont{Browder et~al.}(1997)}]{Browder:1997eu}
\bibinfo{author}{\bibfnamefont{T.~E.} \bibnamefont{Browder}}
  \bibnamefont{et~al.} (\bibinfo{collaboration}{CLEO}), \bibinfo{journal}{Phys.
  Rev.} \textbf{\bibinfo{volume}{D56}}, \bibinfo{pages}{5359}
  (\bibinfo{year}{1997}), \eprint{hep-ex/9706005}.

\bibitem[{\citenamefont{Bergstr{\"o}m}(1982)}]{Bergstrom:1982zq}
\bibinfo{author}{\bibfnamefont{L.}~\bibnamefont{Bergstr{\"o}m}},
  \bibinfo{journal}{Zeit. Phys.} \textbf{\bibinfo{volume}{C14}},
  \bibinfo{pages}{129} (\bibinfo{year}{1982}).

\bibitem[{\citenamefont{Bargholtz et~al.}(2007)}]{Bargholtz:2006gz}
\bibinfo{author}{\bibfnamefont{C.}~\bibnamefont{Bargholtz}}
  \bibnamefont{et~al.} (\bibinfo{collaboration}{CELSIUS/WASA}),
  \bibinfo{journal}{Phys. Lett.} \textbf{\bibinfo{volume}{B644}},
  \bibinfo{pages}{299} (\bibinfo{year}{2007}), \eprint{hep-ex/0609007}.

\bibitem[{\citenamefont{Yao et~al.}(2006)}]{Yao:2006px}
\bibinfo{author}{\bibfnamefont{W.~M.} \bibnamefont{Yao}} \bibnamefont{et~al.}
  (\bibinfo{collaboration}{Particle Data Group}), \bibinfo{journal}{J. Phys.}
  \textbf{\bibinfo{volume}{G33}}, \bibinfo{pages}{1} (\bibinfo{year}{2006}).

\bibitem[{\citenamefont{Jarlskog and Pilkuhn}(1967)}]{Jarlskog:1967aa}
\bibinfo{author}{\bibfnamefont{C.}~\bibnamefont{Jarlskog}} \bibnamefont{and}
  \bibinfo{author}{\bibfnamefont{H.}~\bibnamefont{Pilkuhn}},
  \bibinfo{journal}{Nucl. Phys.} \textbf{\bibinfo{volume}{B1}},
  \bibinfo{pages}{264} (\bibinfo{year}{1967}).

\bibitem[{\citenamefont{Picciotto and Richardson}(1993)}]{Picciotto:1993aa}
\bibinfo{author}{\bibfnamefont{C.}~\bibnamefont{Picciotto}} \bibnamefont{and}
  \bibinfo{author}{\bibfnamefont{S.}~\bibnamefont{Richardson}},
  \bibinfo{journal}{Phys. Rev.} \textbf{\bibinfo{volume}{D48}},
  \bibinfo{pages}{3395} (\bibinfo{year}{1993}).

\bibitem[{\citenamefont{Faessler et~al.}(2000)\citenamefont{Faessler, Fuchs,
  and Krivoruchenko}}]{Faessler:1999de}
\bibinfo{author}{\bibfnamefont{A.}~\bibnamefont{Faessler}},
  \bibinfo{author}{\bibfnamefont{C.}~\bibnamefont{Fuchs}}, \bibnamefont{and}
  \bibinfo{author}{\bibfnamefont{M.~I.} \bibnamefont{Krivoruchenko}},
  \bibinfo{journal}{Phys. Rev.} \textbf{\bibinfo{volume}{C61}},
  \bibinfo{pages}{035206} (\bibinfo{year}{2000}), \eprint{nucl-th/9904024}.

\bibitem[{\citenamefont{Bijnens and Borg}(1999)}]{Bijnens:1999jp}
\bibinfo{author}{\bibfnamefont{J.}~\bibnamefont{Bijnens}} \bibnamefont{and}
  \bibinfo{author}{\bibfnamefont{F.}~\bibnamefont{Borg}}
  (\bibinfo{year}{1999}), \eprint{hep-ph/0106130}.

\bibitem[{\citenamefont{Borasoy and Nissler}(2007)}]{Borasoy:2007dw}
\bibinfo{author}{\bibfnamefont{B.}~\bibnamefont{Borasoy}} \bibnamefont{and}
  \bibinfo{author}{\bibfnamefont{R.}~\bibnamefont{Nissler}},
  \bibinfo{journal}{Eur. Phys. J.} \textbf{\bibinfo{volume}{A33}},
  \bibinfo{pages}{95} (\bibinfo{year}{2007}), \eprint{arXiv:0705.0954
  [hep-ph]}.

\bibitem[{\citenamefont{Lopez et~al.}(2007)}]{Lopez:2007pp}
\bibinfo{author}{\bibfnamefont{A.}~\bibnamefont{Lopez}} \bibnamefont{et~al.}
  (\bibinfo{collaboration}{CLEO}) (\bibinfo{year}{2007}),
  \eprint{arXiv:0707.1601 [hep-ex]}.

\bibitem[{\citenamefont{Drell}(1959)}]{Drell:1959aa}
\bibinfo{author}{\bibfnamefont{S.~D.} \bibnamefont{Drell}},
  \bibinfo{journal}{Nuovo Cimento} \textbf{\bibinfo{volume}{11}},
  \bibinfo{pages}{693} (\bibinfo{year}{1959}).

\bibitem[{\citenamefont{Gomez~Dumm and Pich}(1998)}]{GomezDumm:1998gw}
\bibinfo{author}{\bibfnamefont{D.}~\bibnamefont{Gomez~Dumm}} \bibnamefont{and}
  \bibinfo{author}{\bibfnamefont{A.}~\bibnamefont{Pich}},
  \bibinfo{journal}{Phys. Rev. Lett.} \textbf{\bibinfo{volume}{80}},
  \bibinfo{pages}{4633} (\bibinfo{year}{1998}), \eprint{hep-ph/9801298}.

\bibitem[{\citenamefont{Isidori and Unterdorfer}(2004)}]{Isidori:2003ts}
\bibinfo{author}{\bibfnamefont{G.}~\bibnamefont{Isidori}} \bibnamefont{and}
  \bibinfo{author}{\bibfnamefont{R.}~\bibnamefont{Unterdorfer}},
  \bibinfo{journal}{JHEP} \textbf{\bibinfo{volume}{01}}, \bibinfo{pages}{009}
  (\bibinfo{year}{2004}), \eprint{hep-ph/0311084}.

\bibitem[{\citenamefont{Savage et~al.}(1992)\citenamefont{Savage, Luke, and
  Wise}}]{Savage:1992ac}
\bibinfo{author}{\bibfnamefont{M.~J.} \bibnamefont{Savage}},
  \bibinfo{author}{\bibfnamefont{M.~E.} \bibnamefont{Luke}}, \bibnamefont{and}
  \bibinfo{author}{\bibfnamefont{M.~B.} \bibnamefont{Wise}},
  \bibinfo{journal}{Phys. Lett.} \textbf{\bibinfo{volume}{B291}},
  \bibinfo{pages}{481} (\bibinfo{year}{1992}), \eprint{hep-ph/9207233}.

\bibitem[{\citenamefont{Ametller et~al.}(1993)\citenamefont{Ametller, Bramon,
  and Masso}}]{Ametller:1993we}
\bibinfo{author}{\bibfnamefont{L.}~\bibnamefont{Ametller}},
  \bibinfo{author}{\bibfnamefont{A.}~\bibnamefont{Bramon}}, \bibnamefont{and}
  \bibinfo{author}{\bibfnamefont{E.}~\bibnamefont{Masso}},
  \bibinfo{journal}{Phys. Rev.} \textbf{\bibinfo{volume}{D48}},
  \bibinfo{pages}{3388} (\bibinfo{year}{1993}), \eprint{hep-ph/9302304}.

\bibitem[{\citenamefont{Abouzaid et~al.}(2007)}]{Abouzaid:2006kk}
\bibinfo{author}{\bibfnamefont{E.}~\bibnamefont{Abouzaid}} \bibnamefont{et~al.}
  (\bibinfo{collaboration}{KTeV}), \bibinfo{journal}{Phys. Rev.}
  \textbf{\bibinfo{volume}{D75}}, \bibinfo{pages}{012004}
  (\bibinfo{year}{2007}), \eprint{hep-ex/0610072}.

\bibitem[{\citenamefont{Dorokhov and Ivanov}(2007)}]{Dorokhov:2007bd}
\bibinfo{author}{\bibfnamefont{A.~E.} \bibnamefont{Dorokhov}} \bibnamefont{and}
  \bibinfo{author}{\bibfnamefont{M.~A.} \bibnamefont{Ivanov}},
  \bibinfo{journal}{Phys. Rev.} \textbf{\bibinfo{volume}{D75}},
  \bibinfo{pages}{114007} (\bibinfo{year}{2007}), \eprint{arXiv:0704.3498
  [hep-ph]}.

\bibitem[{\citenamefont{Kahn et~al.}(2007)\citenamefont{Kahn, Schmitt, and
  Tait}}]{Kahn:2007ru}
\bibinfo{author}{\bibfnamefont{Y.}~\bibnamefont{Kahn}},
  \bibinfo{author}{\bibfnamefont{M.}~\bibnamefont{Schmitt}}, \bibnamefont{and}
  \bibinfo{author}{\bibfnamefont{T.}~\bibnamefont{Tait}}
  (\bibinfo{year}{2007}), \eprint{arXiv:0712.0007 [hep-ph]}.

\bibitem[{\citenamefont{Zabierowski et~al.}(2002)}]{Zabierowski:2002ah}
\bibinfo{author}{\bibfnamefont{J.}~\bibnamefont{Zabierowski}}
  \bibnamefont{et~al.} (\bibinfo{collaboration}{CELSIUS/WASA}),
  \bibinfo{journal}{Phys. Scripta} \textbf{\bibinfo{volume}{T99}},
  \bibinfo{pages}{159} (\bibinfo{year}{2002}).

\bibitem[{\citenamefont{Ekstr{\"o}m}(2002)}]{Ekstrom:2002ai}
\bibinfo{author}{\bibfnamefont{C.}~\bibnamefont{Ekstr{\"o}m}}
  (\bibinfo{collaboration}{CELSIUS/WASA}), \bibinfo{journal}{Phys. Scripta}
  \textbf{\bibinfo{volume}{T99}}, \bibinfo{pages}{169} (\bibinfo{year}{2002}).

\bibitem[{\citenamefont{Bargholtz et~al.}(1997)}]{Bargholtz:1997zr}
\bibinfo{author}{\bibfnamefont{C.}~\bibnamefont{Bargholtz}}
  \bibnamefont{et~al.}, \bibinfo{journal}{Nucl. Instrum. Meth.}
  \textbf{\bibinfo{volume}{A390}}, \bibinfo{pages}{160} (\bibinfo{year}{1997}).

\bibitem[{\citenamefont{Feldman and Cousins}(1998)}]{Feldman:1997qc}
\bibinfo{author}{\bibfnamefont{G.~J.} \bibnamefont{Feldman}} \bibnamefont{and}
  \bibinfo{author}{\bibfnamefont{R.~D.} \bibnamefont{Cousins}},
  \bibinfo{journal}{Phys. Rev.} \textbf{\bibinfo{volume}{D57}},
  \bibinfo{pages}{3873} (\bibinfo{year}{1998}), \eprint{physics/9711021}.

\end{thebibliography}
\end{document}